# A simple measure of economic complexity

Sabiou Inoua

*Economic Science Institute, Chapman University, 1 University Dr., Orange, CA 92866, United States of America*

ARTICLE INFO

ABSTRACT

Contrary to conventional economic growth theory, which reduces a country's output to one aggregate variable (GDP), product diversity is central to economic development, as recent "economic complexity" research suggests. A country's product diversity reflects its diversity of knowhow or "capabilities". Researchers proposed the Economic Complexity Index (ECI) and the country Fitness index to estimate a country's number of capabilities from international export data; these measures predict economic growth better than conventional variables such as human capital. This paper offers a simpler measure of a country's knowhow, Log Product Diversity (or LPD, the logarithm of a country's number of products), which can be derived from a one-parameter combinatorial model of production in which a set of knowhows combine with some probability to turn raw materials into a product. ECI and log-fitness can be interpreted theoretically (using the combinatorial model) and empirically as potentially noisy estimates of LPD; moreover, controlling for natural resources, the simple measure better explains the cross-country differences in GDP and in GDP per capita.

## 1. Introduction

The standard view on economic growth and development reduces a country's production to one aggregate variable, GDP; yet economic development is strongly associated with product diversification: rich countries (in terms of GDP per capita) make diverse products, including highly sophisticated ones (e.g., the USA make all products made worldwide: 772 out of 772), whereas poorer countries make only a few rudimentary ones. More generally, product diversity is a good indicator of economic development (Table 1).[1]

The complexity of a country's production (the diversity and sophistication of its products) reflects the country's diversity of productive knowledge, or "capabilities", which combine to transform raw materials into various products, the knowledge content of which ranging in theory from zero for naturally occurring goods (say natural resources sold in the raw) to the maximum value for products requiring all available knowhow (consider an aircraft, e.g.). So, in theory, one can define the knowledge sophistication (or complexity) $S$ of a product by the number of capabilities its production requires; and the knowledge sophistication of a country's whole output, by the total number $K$ of capabilities its production involves. The variables $K$ and $S$ are not directly observable. Researchers proposed the Economic Complexity Index (ECI) and the Product Complexity Index (PCI), respectively, as estimates of $K$ and $S$ computed from international export data, modeled as a network linking countries to the products they make: ECI and PCI are mutually determined through an algorithm akin to the one the web search engine Google uses to rank webpages (Hidalgo and Hausmann, 2009; Hausmann and Hidalgo, 2014).[2] The ECI, as its authors showed, explains economic growth better than conventional variables such as human capital (Hausmann and Hidalgo, 2014, Section 4). Alternative complexity measures include the country Fitness and product Quality indices (Tacchella et al., 2012; Cristelli et al., 2013).

Here I propose a simpler measure of a country's knowhow, Log Product Diversity (LPD), which outperforms the network metrics on at least three grounds. First, it is theoretically derived, from a one-parameter combinatorial model of production that is more directly in line with an early metaphor of the production process as a game of scrabble (Hausmann and Hidalgo, 2014, p. 20) and that is entirely based on the basic assumption that a set of capabilities combine with some probability to transform raw materials into a product. The model's core implication follows immediately if one assumes that any random combination of knowhows is a meaningful knowhow: then with $K$ elementary knowhows (or capabilities), a country can make up to $D = 2^K$ products with sophistications $S = 0, 1, 2, \ldots, K$. Thus, one can estimate $K$

*E-mail address:* inoua@chapman.edu.

[1] The data used will be described in Section 3.

[2] See Eqs. (11)–(14) below for a stylized reminder on the definitions of ECI and Fitness, and the original papers, or this paper's Appendix A, for the technical details involved in their derivations.

**Table 1**
The world's most and least product-diversified economies (2018).[a]

| Ten most diversified economies | | Ten least diversified economies | |
|---|---|---|---|
| *Country* | *Diversification* | *Country* | *Diversification* |
| USA | 772 | Liberia | 249 |
| Germany | 771 | Burundi | 227 |
| Spain | 770 | Cuba | 217 |
| UK | 770 | Haiti | 216 |
| China | 769 | Yemen | 213 |
| Italy | 769 | Libya | 189 |
| Netherlands | 768 | Djibouti | 178 |
| South Africa | 768 | Somalia | 141 |
| France | 767 | Central Afr. Rep. | 123 |
| India | 766 | Chad | 120 |

[a] Small islands excluded, these being among the least diversified economies. More on the peculiarity of island economies in Section 4.

by $\log(D)$ up to a scaling constant [the inverse of $\log(2)$]. More realistically, assume that a combination of knowhows is productively relevant (can be used to turn a raw material) only with some probability: then *log* ($D$) is still an estimate of $K$, but up to a scaling factor that can vary across countries due to their different endowments in raw materials or natural resources more generally. This is the essence of the combinatorial model of economic development.

The combinatorial model differs by its simplicity from previous able ones (e.g., Hausmann and Hidalgo, 2011). Its natural language is informational in the sense of Shannon (1948), which follows from the definition of a capability as the smallest unit of information needed to encode the knowledge required to make a product. Granted this more general informational interpretation, the model offers in fact a unified theoretical framework for the complex view on economic development; in particular, one can show from it that both ECI and log-fitness are estimates of LPD, if potentially noisy ones due to the interdependent nature of the two network metrics (according to which a country's estimated knowhow depends mutually on that of similar countries and of their products, so that an error in one estimate propagates to others; whereas the simple measure is self-contained and hence free from this problem).[3]

Besides its more explicit grounding on theory (derived in Section 2) and its simple data requirement (explained in Section 3), LPD explains GDP and GDP per capita better than ECI and Fitness (as argued in Section 4).

## 2. Model and predictions

The theoretical model is centered on one assumption, which I call *the combinatorial hypothesis*: capabilities combine with some probability $\pi$ to form a coherent set of knowhow; more precisely, the probability that any combination of $S$ capabilities makes sense as a coherent set of productive knowhows (that can be used to transform raw materials into a product) decays exponentially with $S$ at a rate $\pi$. Merely to minimize the mathematical symbolism, I assume further that every country possesses (with probability 1) the set of raw materials to transform with its capabilities. The more general formulation would amount in effect to treating $\pi$ as a joint probability involving the probability of finding the raw materials to transform (making $\pi$ variable across countries and products). It will prove convenient to collectively name the various empirical departures from the auxiliary assumption as the *natural-resource-bias* (NRB), which pertains more generally to naturally occurring products (in practice, very rudimentary products sold almost in the raw) and their exporters.

The probability of a product requiring $S$ capabilities (or $S$-product for short) is by assumption

$$\text{prob}(S\text{-product}) = \pi^S. \tag{1}$$

A country that has $K$ capabilities is expected to make a number of products given by the binomial formula[4]:

$$D = \sum_{S=0}^{K} \binom{K}{S} \pi^S = (1+\pi)^K. \tag{2}$$

Thus, in this model, an economy (or country) $c$ endowed with $K_c$ capabilities is entirely characterized by its product list $\{1, \ldots, p, \ldots, D_c\}$, where $D_c = (1+\pi)^{K_c}$. Observing that a country $c$ makes $D_c$ products, I can infer from (2) that the country has (up to a scaling constant) a number of capabilities given by

$$K_c = \log D_c. \tag{3}$$

By Assumption (1), the complexity or sophistication of a product $p$ is (up to scaling)[5]:

$$S_p \equiv -\log \text{prob}(p). \tag{4}$$

Assumption (1) can in fact be interpreted as a definition or identification of the elementary units of knowhow (capabilities) as the smallest units of information needed to encode the knowledge needed to make a product. From this informational identification of product sophistication $S$ derives therefore an empirical measure of $S$ given by the log-frequency of a product in the world economy, which is modeled theoretically as a population of economies numbered $c = 1, \ldots, C$, each specified by its product list $\{1, \ldots, D_c\}$, and collectively making the product types $\{1, \ldots, P\}$. Now, fix a product $p$ in $\{1, \ldots, P\}$ whose sophistication $S_p$ is to be estimated. By the law of total probability, the frequency of $p$ in the world economy (all countries considered) is a sum whose generic term is the frequency of $p$ among the products of country $c$ multiplied by the frequency of such country in the world economy:

$$\text{prob}(p) = \sum_{c=1}^{C} \text{prob}(p|c)\text{prob}(c), \tag{5}$$

where $\text{prob}(p|c)$ is the frequency of $p$ in country $c$ and $\text{prob}(c)$, the frequency of country $c$ in the world economy. The simplest empirical estimate for $\text{prob}(p)$, which I denote $\Pi_p$, is obtained by setting $\text{prob}(c) = 1/C$ and $\text{prob}(p|c) = M_{cp}/D_c$ in (5), where

$$M_{cp} = \begin{cases} 1, \text{if country } c \text{ makes product } p, \\ 0, \text{otherwise}, \end{cases} \tag{6}$$

namely the famous country-product binary matrix,

$$D_c = \sum_{p=1}^{P} M_{cp}, \tag{7}$$

and $C$ and $P$ being respectively the number of countries and products in the world economy. Then the general theoretical formula (5) specializes into:

$$\Pi_p = \sum_{c=1}^{C} \frac{M_{cp}}{D_c} \frac{1}{C}. \tag{8}$$

In summary, given the matrix $[M_{cp}]$, I estimate $K_c$ and $S_p$ respectively by the informational measures $\log(D_c)$ and $-\log(\Pi_p)$, which I refer to as Log Product Diversity (LPD) and Log Product Probability (LPP):

$$LPD_c = \log(D_c), \tag{9}$$

$$LPP_p = \log \frac{C}{\sum_{c=1}^{C} \frac{M_{cp}}{D_c}}. \tag{10}$$

[3] I thank a reviewer for pointing this out.

[4] As a reviewer pointed out, one may consider investigating how the model's predictions would be altered were the order of capabilities in each combination to matter.

[5] A three-bar equality denotes a definition or an identity.

In contrast to these simple, theoretically derived, measures of knowhow, the algorithmically computed complexity measures, ECI-PCI and Fitness-Quality, are mutually determined in a circular manner, based on the intuition that an economy's total knowhow content is reflected in its products, and, vice versa, a product reflects the average knowhow of its producers; more precisely, the four complexity measures are (up to scaling) solutions to the equations[6]:

$$ECI_c = \sum_p \left(\frac{M_{cp}}{\sum_p M_{cp}}\right) PCI_p, \tag{11}$$

$$PCI_p = \sum_c \left(\frac{M_{cp}}{\sum_c M_{cp}}\right) ECI_c, \tag{12}$$

$$F_c = \sum_p M_{cp} Q_p, \tag{13}$$

$$Q_p = \left[\sum_c M_{cp} F_c^{-1}\right]^{-1}. \tag{14}$$

The combinatorial model of production, as it turns out, does not just provide simpler knowhow measures, but also sheds light on simple principles underlying the alternative measures. For example, the ECI-PCI algorithm assumes that a country knowhow content can be estimated by the country's average product knowhow content, which is indeed true theoretically [Appendix A.1]:

$$\mathbf{E}(S|K) \equiv \sum_{S=0}^{K} S\mathrm{prob}(S|K) = \frac{\pi}{1+\pi} K, \tag{15}$$

where $\mathbf{E}$ stands for mathematical expectation, $\mathbf{E}(S|K)$ the conditional expectation of product sophistication $S$ in a country with $K$ capabilities (or $K$-country for short), and $\mathrm{prob}(S|K)$ the theoretical distribution of $S$ (the proportion of $S$-products) in a $K$-country. PCI is by construction an estimate of $S$, meaning the two variables are equal up to scaling, potential natural-resource-bias (NRB), and potential estimation error:

$$PCI = S + NRB + \mathrm{error}. \tag{16}$$

Thus, for a $K$-country, ECI is expected to estimate the following theoretical value (up to scaling):

$$\mathbf{E}(PCI|K) = K + \mathbf{E}(NRB|K) + \mathbf{E}(\mathrm{error}|K).$$

All in all, the model predicts therefore (up to scaling):

$$ECI_c = LPD_c + NRB_c + \mathrm{error}_c, \tag{17}$$

where the error term depends on the extent to which the theoretical probability $\mathrm{prob}(S|c)$ can be estimated by the frequency of each product $p$ in country $c$, or $M_{cp}/\sum_p M_{cp}$.

Similarly, the model also provides a theoretical rationale for the Fitness-Quality algorithm. Define the average product quality in country $c$ and the average product quality of the producers of product $p$ in terms of a (weighted) arithmetic and harmonic mean, respectively:

$$\bar{Q}_c \equiv \frac{\sum_p M_{cp} Q_p}{\sum_p M_{cp}}, \tag{18}$$

$$\bar{Q}_p \equiv \frac{\sum_c \frac{M_{cp}}{D_c}}{\sum_c \frac{M_{cp}}{D_c} \frac{1}{Q_c}}. \tag{19}$$

Then by construction [after comparing (9)–(10) and (13)–(14)] it follows:

$$F_c \equiv D_c \bar{Q}_c, \tag{20}$$

$$Q_p \equiv \left(C\Pi_p\right)^{-1} \bar{Q}_p. \tag{21}$$

Taking logs in (20)–(21), we see that Fitness and Quality produce the same complexity measures as LPD and LPP up to scaling and correcting terms:

$$\log(F_c) \equiv \log(D_c) + \log(\bar{Q}_c), \tag{22}$$

$$\log\left(Q_p\right) \equiv -\log\left(\Pi_p\right) + \log\left(\bar{Q}_p\right). \tag{23}$$

The correcting factor $\bar{Q}_p$ in (21) is the distinguishing feature of the Fitness-Quality algorithm and is mostly determined by low-complexity economies or their low-complexity products such as raw materials (Tacchella et al., 2012; Cristelli et al., 2013). It is, in other words, a natural-resource-bias correction term (that is better considered in log). Thus (23) reads

$$\log\left(Q_p\right) = LPP_p + NRB_p. \tag{24}$$

By construction, the probability $\Pi_p$ is an estimate of $\mathrm{prob}(p)$, the true probability of observing product $p$ in the world economy, which under the combinatorial hypothesis is $\pi^{S_p}$. Thus up to scaling

$$\log\left(\Pi_p\right) = -S_p - \mathrm{error}_p. \tag{25}$$

So (24) becomes

$$\log\left(Q_p\right) = S_p + NRB_p + \mathrm{error}_p. \tag{26}$$

Neglecting the natural-resource-bias and error terms in (26) yields the theoretical simplification $\log(Q) = S$ (up to scaling), which according to the combinatorial model [Appendix A.2] implies that (up to scaling)

$$\mathbf{E}(Q|K) = D^{-1} 2^K. \tag{27}$$

Hence, for a K-country, Fitness is expected to estimate the theoretical value

$$\mathbf{E}(DQ|K) = 2^K. \tag{28}$$

Theoretically, therefore, log($F$) is proportional to $K$, which is itself proportional to log($D$), so that, all in all, (22) simplifies (up to scaling) to

$$\log(F_c) = \log(D_c) + NRB_c + \mathrm{error}_c. \tag{29}$$

In summary, the predictions (17), (24), and (29), which relate the complexity measures and read more precisely in standardized form (to avoid dealing with scaling constants), come down to saying that, up to natural-resource bias and estimation error:

$$\frac{LPD_c - \mathrm{mean}(LPD)}{\mathrm{std}(LPD)} = ECI_c = \frac{\log(F_c) - \mathrm{mean}(\log(F))}{\mathrm{std}(\log(F))}, \tag{30}$$

$$\frac{LPP_p - \mathrm{mean}(LPP)}{\mathrm{std}(LPP)} = PCI_p = \frac{\log\left(Q_p\right) - \mathrm{mean}(\log(Q))}{\mathrm{std}(\log(Q))}. \tag{31}$$

## 3. Data and method

The whole analysis, recall, presupposes simple data: for any country, the list of products it makes, formally, the country-product binary matrix $\mathbf{M} = [M_{cp}]$ connecting countries to the products they make: $M_{cp} = 1$ if country $c$ makes product $p$, and $M_{cp} = 0$, otherwise. The data should be sufficiently disaggregated in terms of number of products and, for meaningful international comparisons, the products should be listed under a unified nomenclature such as the Standard International Trade Classification (SITC). Simple as they are, however, such data are not directly available; hence economic complexity researchers commonly used as a proxy for a country's product list, its export list as obtained from UN Comtrade (Commodities Trade Statistics database), which for our purpose amounts to the export matrix $\mathbf{X} = [X_{cp}]$, where $X_{cp}$ is the amount country $c$ exported in good $p$. While in principle there will inevitably be some error in approximating a country's product composition by its export mix, the error proves small nonetheless a posteriori, given the accuracy of the results (apparently, countries' exported products are representative of their total product compositions).

[6] For the technical details, see the original works (Hidalgo and Hausmann, 2009; Hausmann and Hidalgo, 2014; Tacchella et al., 2012) or Appendix A.3 of this paper. The economic complexity literature has expanded enormously since the first draft of this paper has been written; because I am going back to the fundamentals of this fascinating field of research, I do not attempt here to address or reflect some of the most recent trends in that literature; for reviews, see, e.g., Hidalgo (2021) and Balland et al. (2022).

The results presented throughout this paper are based on the following matrix:

$$M_{cp} = \begin{cases} 1, & \text{if } X_{cp} > 0, \\ 0, & \text{if } X_{cp} = 0, \end{cases} \tag{32}$$

using the UN Comtrade data under SITC (revision 2).[7] In contrast, ECI and Fitness are based on a country's most competitive exports, meaning those products in which the country has a revealed comparative advantage (RCA) greater than 1, leading to a more restrictive definition of the matrix $\mathbf{M}$ thus

$$M_{cp} = \begin{cases} 1, & \text{if } RCA_{cp} \geq 1, \\ 0, & \text{if } RCA_{cp} < 1, \end{cases} \tag{33}$$

where $RCA_{cp}$ is the revealed comparative advantage of country $c$ in product $p$ (Balassa, 1965), defined as

$$RCA_{cp} = \left( X_{cp} \Big/ \sum_p X_{cp} \right) \Big/ \left( \sum_c X_{cp} \Big/ \sum_{cp} X_{cp} \right).$$

The income data used below are GDP data (in purchasing power parity) from the Penn World Table (PWT8); the preponderance of natural resources in an economy is measured by the share of natural resource export rents (NRR) in GDP or GDP per capita (GDPpc) using data from the World Bank database.[8]

I first test the predicted relationships between the six complexity measures and then compare empirically the country measures (LPD, ECI, and Fitness) by their capacity of explaining GDP and GDP per capita (GDPpc). I use, to this effect, rank regression involving GDP or GDPpc as dependent variable and each one of the three production complexity measures as independent variable and control for natural resources. Rank regression (as opposed to simple regression) is needed because only rank transforms yield a linear correlation between GDP or GDPpc and LPD, which, as will become clear below (Fig. 3), exhibits a near saturation level (achieved by countries such as the US that produce, or export, almost all product types), leading to a nonlinear correlation with GDP or GDPpc; because the other measures present no such nonlinearity, direct comparison of linear correlations (or regressions) of the three economic complexity measures with GDP or GDPpc is not meaningful; in contrast, comparison of rank correlations or regressions is perfectly legitimate.[9]

## 4. Results and discussions

The theoretical predictions (30) and (31) relating the six complexity measures are strongly supported by the data (Figs. 1 and 2). The natural-resource bias reveals itself strikingly [e.g. in Subplot 3 of Fig. 1 relating ECI and LPD] through a tripartition of countries by their differentiated regimes vis-à-vis the theoretical predictions, a tripartition that holds throughout the analysis (the legend of Fig. 1, displayed in Subplot 2, is assumed in all illustrations hereafter): island economies (mostly outliers) exporting essentially raw products (naturally occurring goods); countries exporting (more refined) raw materials such as oil; and the rest of the word. Expectedly, the model's predictions are more accurate if one considers countries' complete export lists (rather than just their most RCA-competitive exports) as representative of their product lists. To compare the three complexity measures as development determinants, however, I confine the discussion from now on to the original definition of ECI and Fitness (assuming only RCA-exports) but keep my definition of LPD (including all exports).

That the nontrivial predictions (30)–(31) are strongly supported by the data is a good support for the plausibility of their assumptions, which, recall, are: (i) the combinatorial hypothesis (the model) holds; (ii) $\Pi_p$ is an estimate of $\text{prob}(p)$; and (iii) PCI is a good estimate of $S$. In other words, the strong empirical support for (30)–(31) is a good indication for the plausibility of the combinatorial hypothesis and the accuracy, up to small noisy errors and the natural-resource bias, of the six complexity metrics as measures of the number of capabilities in a country or required by a product. It remains to analyze and compare the relevance of the economic complexity measures for economic development.

LPD is strongly correlated with GDP for most countries (with small islands and natural-resource exporters appearing as outliers as expected: Fig. 3, Top Subplot) but in a nonlinear way due to the saturation effect of highly diversified economies (which make nearly all products and hence have nearly the same $D$).

Some countries of course owe their riches to a few natural resources—notably oil—and tend to have higher incomes given their product diversity; in compensation, one would expect the output in these countries is more volatile. LPD, along with natural resource-rent (NRR), explains much of GDP in terms of rank regression and does so better than ECI and Fitness. In essence countries' GDP ranking is a weighted average of their economic complexity (LPD) and natural resource (NRR) rankings (Table 2).

The link between economic complexity and economic development is strongest when the former is measured by LPD and becomes particularly interesting if the latter is measured in terms of GDP per capita (GDPpc), for then the analysis shows a clear picture of known ideas such as the so-called "natural resource curse" (Frankel, 2010; Badeeb et al., 2017). Thus, a country's GDPpc is dominantly determined by its knowhow similarly to GDP; but after controlling for knowhow, dependence on natural resources (which I measure now by the fraction of natural resource export rents in GDPpc, still denoting it abusively as NRR) now contributes negatively to GDPpc (Table 3); put the other way around, natural-resource-intensity adds to a country's overall income but not to the average welfare of its population after controlling for knowhow.

The regression results, be it insisted, hold after controlling for natural resources, in accordance with the combinatorial model of production; only absent this natural-resource control is LPD outperformed by ECI in explaining GDPpc (Table 5), while reamining the better explanator of GDP even without the control (Table 4).

[7] For the raw data, see https://comtrade.un.org/Data/. I used the data as corrected for discrepancies by the Growth Lab at Harvard University ["Growth Lab (2019). International Trade Data (SITC, Rev. 2)"]. These corrections are needed for greater accuracy, not so much of log-product-diversity, but of ECI and Fitness, which are sensitive to errors in the export values.

[8] The PWT is accessible through the GGDC (Groningen Growth and Development Centre, University of Groningen), at https://www.rug.nl/ggdc/productivity/pwt/. I use the RGDPO variable (an output-oriented GDP estimate); but the other measures would give very similar results. The natural-resource-rent data are available at https://data.worldbank.org/indicator/NY.GDP.TOTL.RT.ZS.

[9] Rank regression (consisting of replacing each variable by its statistical rank) is a simple but powerful regression technique especially for nonlinear relationships (see, e.g., Iman and Conover, 1979).

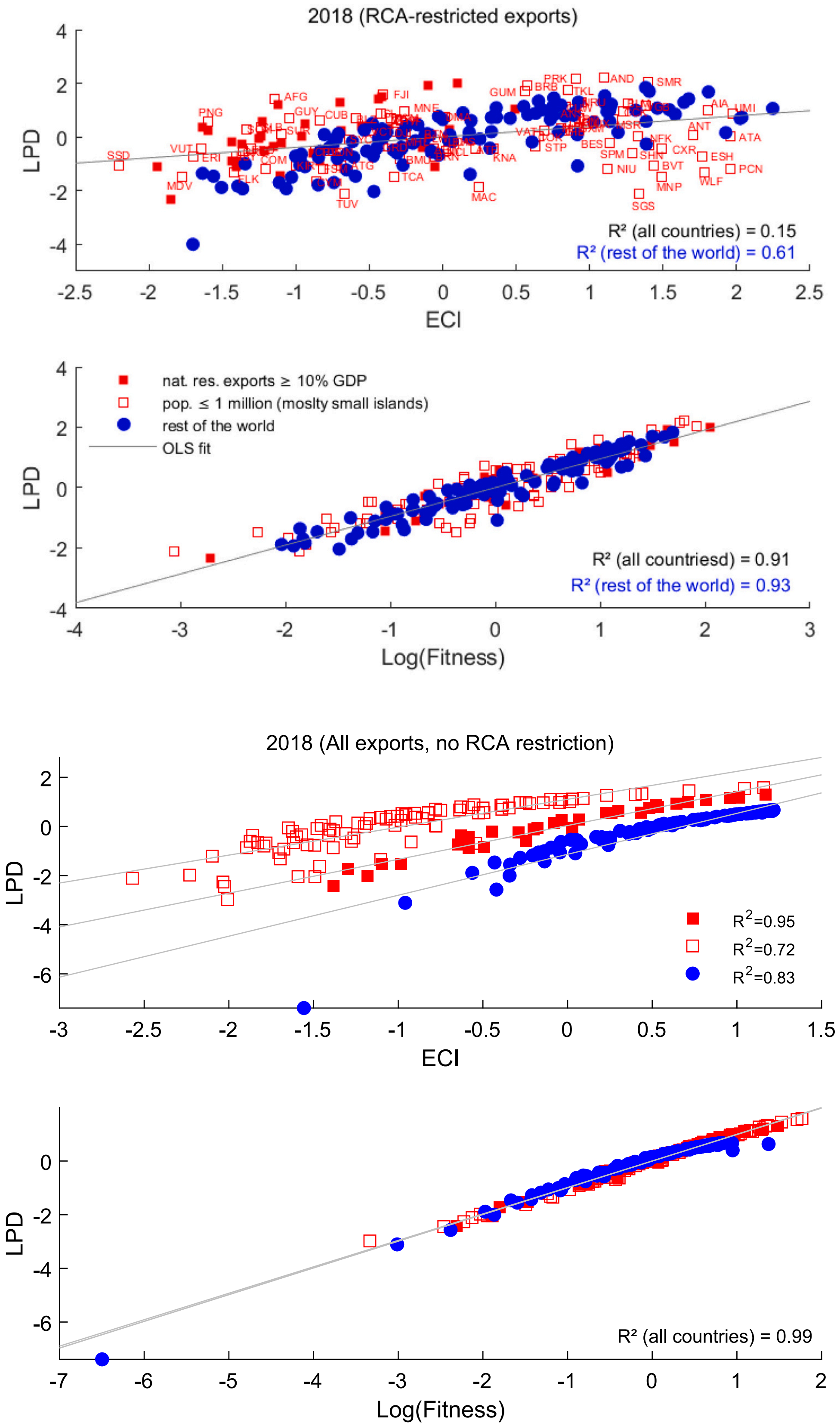

**Fig. 1.** The predicted relationships between the three economic complexity measures (all standardized) are supported by the empirical data. Top panel: RCA-restricted export data. Bottom panel: all exported products. (The solid lines refer to OLS regression fits.)

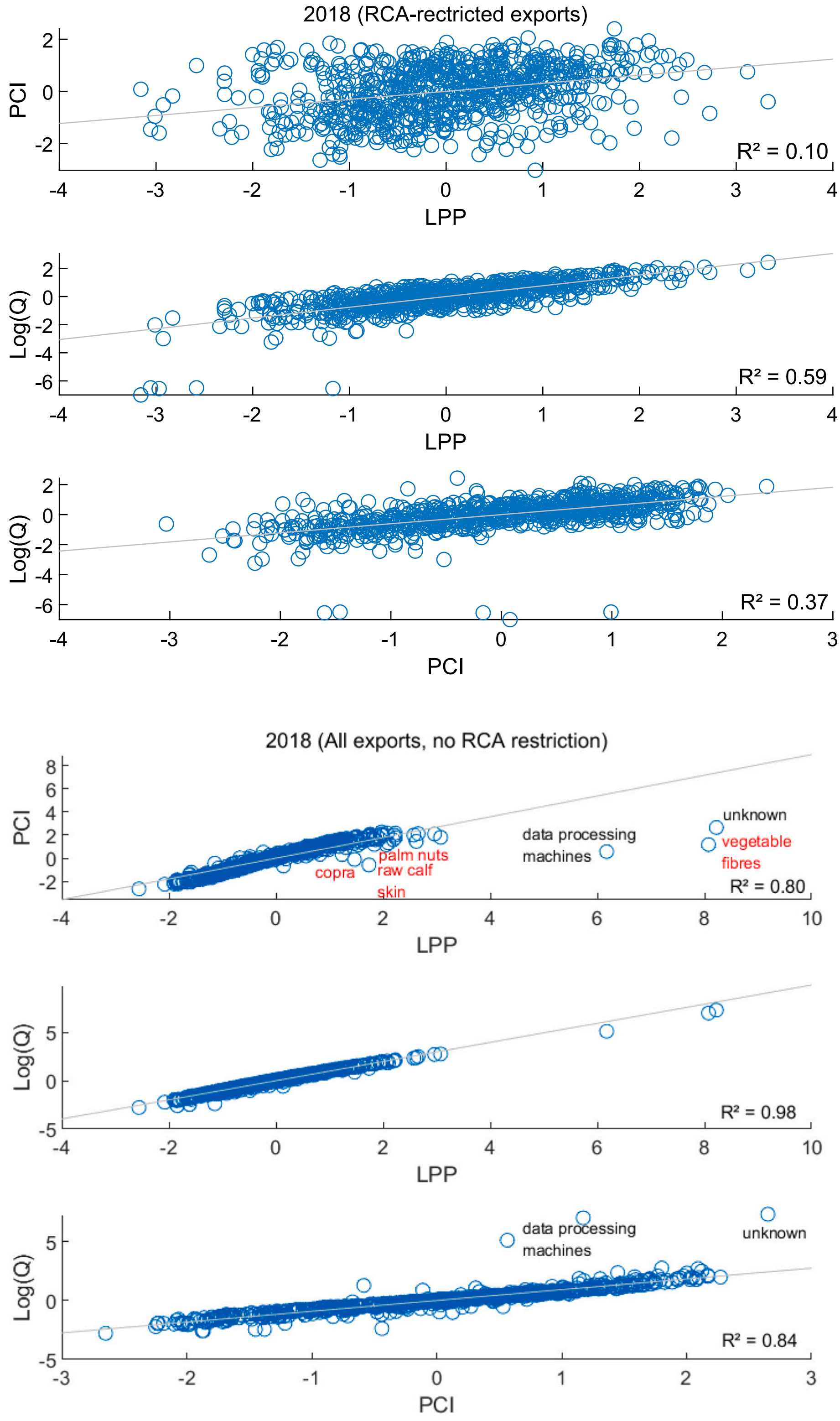

**Fig. 2.** The predicted relationships between the three product complexity measures (all standardized) are also supported by the data. Top: RCA-exports. Bottom: all exports.

## 5. Conclusion

In this paper I lay the foundation for an informational theory of economic development centered on a combinatorial model of production that offers a simple, unifying, theoretical framework for the fast-growing complex view on economic development. From this model derives a simple measure of a country's knowhow, LPD, that is easily computed, requires minimum data, allows a clearer and more directly interpretable perspective on the link between natural resources, country knowhow, and economic development, and moreover explains economic development better than the more commonly used measures, ECI and (log) Fitness, after controlling for natural resources.

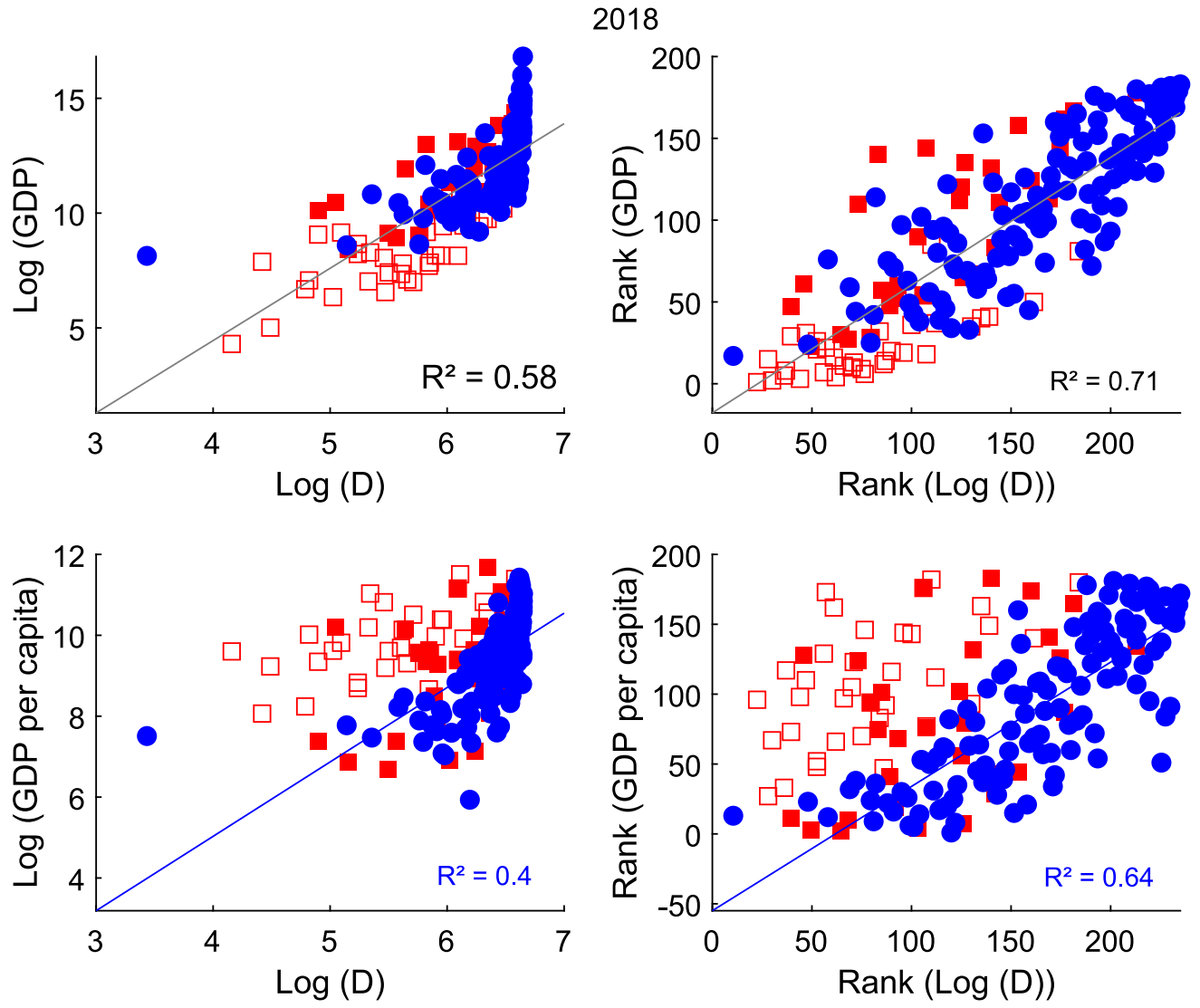

**Fig. 3.** LPD versus GDP and GDP per capita: linear (left) versus rank correlations (right).

**Table 2**
Rank regression results: GPD versus Natural Resource Rent per GDP (NRR) and LPD (top), ECI (middle), and Fitness (bottom).

| | Coef. Est. | S.E. | t-stat. | *p*-value |
|---|---|---|---|---|
| Intercept | −47.56 | 7.05 | −6.75 | $2.22 \times 10^{-10}$ |
| Rank (LPD) | 0.81 | 0.03 | 23.25 | $5.64 \times 10^{-55}$ |
| Rank (NRR) | 0.24 | 0.04 | 6.67 | $3.42 \times 10^{-10}$ |
| $R^2 = 0.76$; number of observations: 175. | | | | |
| Intercept | −20.10 | 12.71 | −1.58 | 0.12 |
| Rank (ECI) | 0.58 | 0.06 | 9.68 | $5.76 \times 10^{-18}$ |
| Rank (NRR) | 0.45 | 0.07 | 6.62 | $4.29 \times 10^{-10}$ |
| $R^2 = 0.36$; number of observations: 175. | | | | |
| Intercept | −12.08 | 8.87 | −1.36 | 0.18 |
| Rank (Fitness) | 0.58 | 0.04 | 14.29 | $4.94 \times 10^{-31}$ |
| Rank (NRR) | 0.27 | 0.05 | 5.43 | $1.92 \times 10^{-7}$ |
| $R^2 = 0.55$; number of observations: 175. | | | | |

**Table 3**
Rank regression results: GDP per capita versus Output Complexity and Natural Resource Rent.

| | Coef. Est. | S.E. | t-stat. | p-value |
|---|---|---|---|---|
| Intercept | 55.4 | 8.38 | 6.61 | $4.64 \times 10^{-10}$ |
| Rank (LPD) | 0.57 | 0.05 | 11.57 | $2.92 \times 10^{-23}$ |
| Rank (NRR) | −0.52 | 0.06 | −9.37 | $3.97 \times 10^{-17}$ |
| $R^2 = 0.53$; number of observations: 175. | | | | |
| Intercept | 57.24 | 9.56 | 5.98 | $1.20 \times 10^{-8}$ |
| Rank (ECI) | 0.48 | 0.05 | 9.52 | $1.57 \times 10^{-17}$ |
| Rank (NRR) | −0.22 | 0.06 | −3.55 | $4.91 \times 10^{-4}$ |
| $R^2 = 0.45$; number of observations: 175. | | | | |
| Intercept | 92.70 | 9.14 | 10.14 | $3.05 \times 10^{-19}$ |
| Rank (Fitness) | 0.29 | 0.05 | 5.78 | $3.45 \times 10^{-8}$ |
| Rank (NRR) | −0.44 | 0.07 | −6.68 | $3.16 \times 10^{-10}$ |
| $R^2 = 0.30$; number of observations: 175. | | | | |

**Table 4**
Rank regression results without NRR control: GDP versus LPD (top), ECI (middle), and Fitness (bottom).

| | Coef. Est. | S.E. | t-stat. | p-value |
|---|---|---|---|---|
| Intercept | −17.80 | 5.62 | −3.17 | $1.80 \times 10^{-3}$ |
| Rank (LPD) | 0.78 | 0.04 | 21.08 | $1.28 \times 10^{-50}$ |
| $R^2 = 0.71$; number of observations: 183. | | | | |
| Intercept | 55.40 | 7.44 | 7.44 | $3.85 \times 10^{-12}$ |
| Rank (ECI) | 0.32 | 0.06 | 5.63 | $6.82 \times 10^{-8}$ |
| $R^2 = 0.15$; number of observations: 175. | | | | |
| Intercept | 22.54 | 6.40 | 3.52 | $5.44 \times 10^{-4}$ |
| Rank (Fitness) | 0.53 | 0.04 | 12.19 | $2.49 \times 10^{-25}$ |
| $R^2 = 0.45$; number of observations: 175. | | | | |

**Table 5**
Rank regression results without NRR control: GDP per capita versus LPD (top), ECI (middle), and Fitness (bottom).

| | Coef. Est. | S.E. | t-stat. | p-value |
|---|---|---|---|---|
| Intercept | 24.11 | 8.92 | 2.70 | $7.50 \times 10^{-3}$ |
| Rank (LPD) | 0.48 | 0.06 | 8.21 | $3.90 \times 10^{-14}$ |
| $R^2 = 0.27$; number of observations: 183. | | | | |
| Intercept | 31.77 | 6.23 | 5.10 | $8.61 \times 10^{-7}$ |
| Rank (ECI) | 0.53 | 0.05 | 11.06 | $4.60 \times 10^{-22}$ |
| $R^2 = 0.40$; number of observations: 175. | | | | |
| Intercept | 56.46 | 8.11 | 6.96 | $5.98 \times 10^{-11}$ |
| Rank (Fitness) | 0.27 | 0.06 | 4.92 | $1.93 \times 10^{-6}$ |
| $R^2 = 0.12$; number of observations: 175. | | | | |

By its simplicity, LPD might also be useful as a clarifying concept in innovation studies because it implicitly derives from a framework that distinguishes two forms of knowhow that are so far mingled in the economic complexity literature: productive knowhow (the skills needed to create products) versus competitive knowhow (the skills needed to competitively sell or export products); strictly speaking, LPD (at least when extended to all products) estimates the former, whereas ECI and Fitness estimate the latter. In fact, there are interesting aspects of the world distribution of knowhow that become evident only after this distinction has been made; for example, one can show that the world distribution of knowhow as measured by LPD has a precise mathematical form: it is a Beta distribution. I intend to discuss this and other unexplored aspects of the model in a follow-up work.

## CRediT authorship contribution statement

The author (Sabiou Inoua) confirms sole responsibility for the article: theoretical model, results, analysis, etc.

## Declaration of competing interest

The author declares that he has no known competing financial interests or personal relationships that could have appeared to influence the work reported in this paper.

## Data availability

Data will be made available on request.

**Acknowledgments**

I thank two anonymous reviewers for their constructive comments and suggestions that significantly improved the paper. Following a suggestion by one of the reviewers, I have reverted to an earlier, less mathematically oriented, version of the paper (Inoua, 2016) that better fits this Journal (and will present in a follow-up work, technical developments herein excluded).

## Appendix A. Technicalities

### A.1. Proof that the average of S given K is proportional to K theoretically

According to the model, recall, a country with $K$ capabilities (a $K$-country for short) makes $(1+\pi)^K$ products among which $\binom{K}{S}\pi^S$ have sophistication $S$. Thus, the distribution of product sophistication in such country is

$$\mathrm{prob}(S|K) = \frac{\binom{K}{S}\pi^S}{(1+\pi)^K}, \quad S = 0, \dots, K. \tag{34}$$

The average product sophistication in a $K$-country is:

$$\mathbf{E}(S|K) \equiv \sum_{S=0}^{K} S\,\mathrm{prob}(S|K) \qquad \text{(by definition)}$$

$$= D^{-1}\sum_{S=1}^{K} S\binom{K}{S}\pi^S$$

$$= D^{-1}\sum_{S=1}^{K} S\frac{K}{S}\binom{K-1}{S-1}\pi^S \qquad \text{(by a known identity)}$$

$$= D^{-1}\pi K\sum_{S=1}^{K}\binom{K-1}{S-1}\pi^{S-1}$$

$$= D^{-1}\pi K\sum_{S'=0}^{K-1}\binom{K-1}{S'}\pi^{S'} \qquad \text{(by setting } S' = S-1)$$

$$= D^{-1}\pi K(1+\pi)^{K-1}.$$

$$= \frac{\pi}{1+\pi}K.$$

### A.2. Proof that log(D) and log(F) are proportional theoretically

Neglecting the natural-resource bias and the estimation error term, the theoretical prediction (26) reads $\log(Q) = S$ (up to scaling) or more precisely

$$Q = \pi^{-S}. \tag{35}$$

Thus, the model predicts that the average product quality in a $K$-country is

$$\mathbf{E}(Q|K) \equiv \sum_{S=0}^{K} Q\,\mathrm{prob}(S|K)$$

$$= \sum_{S=0}^{K}\pi^{-S}\mathrm{prob}(S|K)$$

$$= D^{-1}\sum_{S=0}^{K}\pi^{-S}\binom{K}{S}\pi^S$$

$$= D^{-1}\sum_{S=0}^{K}\binom{K}{S}$$

$$= D^{-1}\sum_{S=0}^{K}\binom{K}{S}$$

$$= D^{-1}2^K.$$

Hence (up to scaling) Fitness (defined as a country's total product quality, or the average product quality times the number of products) is theoretically:

$$F = D\mathbf{E}(Q|K) = 2^K. \tag{36}$$

Since theoretically $D = (1+\pi)^K$, it follows that

$$\frac{\log(F)}{\log(D)} = \frac{\log(2)}{\log(1+\pi)}. \tag{37}$$

### A.3. The ECI-PCI and F-Q algorithms

According to the first algorithm, a country's economic complexity $x$ is proportional to the average complexity of its products and, vice versa, a product's complexity $y$ is proportional to the average complexity of its producers:

$$x_c = \alpha \sum_p W_{cp} y_p, \tag{38}$$

$$y_p = \beta \sum_p W^*_{pc} x_c, \tag{39}$$

where $\alpha$ and $\beta$ are positive scaling constants and the weights

$$W_{cp} \equiv \frac{M_{cp}}{\sum_p M_{cp}}, \tag{40}$$

$$W^*_{pc} \equiv \frac{M_{cp}}{\sum_c M_{pc}}. \tag{41}$$

Collecting the variables and weights into the vectors and matrices $\mathbf{x} = [x_c]$, $\mathbf{y} = \left[y_p\right]$, $\mathbf{W} = \left[W_{cp}\right]$, and $\mathbf{W}^* = \left[W^*_{pc}\right]$, then (38) and (39) become $\mathbf{x} = \alpha\mathbf{W}\mathbf{y}$ and $\mathbf{y} = \beta\mathbf{W}^*\mathbf{x}$, which together yield:

$$(\mathbf{W}\mathbf{W}^*)\mathbf{x} = (\alpha\beta)^{-1}\mathbf{x}, \tag{42}$$

$$(\mathbf{W}^*\mathbf{W})\mathbf{y} = (\alpha\beta)^{-1}\mathbf{y}. \tag{43}$$

That is, economic or product complexity is measured by an eigenvector of $\mathbf{W}\mathbf{W}^*$ and $\mathbf{W}^*\mathbf{W}$, respectively, where the associated eigenvalue is $(\alpha\beta)^{-1}$. Because the averaging weights sum to 1, it is easy to see that any (positive) uniform vectors are solutions to this eigenvector problem; these are the eigenvectors associated with the largest eigenvalue, which can be shown to be 1. Thus, the authors of this algorithm choose the eigenvectors associated with the *second largest eigenvalue*. Finally, the ECI and PCI are (up to the sign) the elements of the chosen eigenvectors given in standardized form:

$$\text{ECI} = +\,\text{sign}[\text{corr}(\mathbf{x}, \mathbf{d})]\,\frac{\mathbf{x} - \text{mean}(\mathbf{x})}{\text{std}(\mathbf{x})}, \tag{44}$$

$$\text{PCI} = -\,\text{sign}[\text{corr}(\mathbf{y}, \mathbf{u})]\,\frac{\mathbf{y} - \text{mean}(\mathbf{y})}{\text{std}(\mathbf{y})}, \tag{45}$$

where $\mathbf{d}$ and $\mathbf{u}$ are the vectors of country product diversity and product so-called ubiquity (the number of countries making a product): $\mathbf{d} \equiv [D_c]$ and $\mathbf{u} \equiv \left[U_p\right]$, with $U_p \equiv \sum_c M_{cp}$. I multiplied the standardized eigenvectors with the signed correlations to ensure the signs of ECI and PCI are correct (product complexity is inversely correlated with product ubiquity); this is simply because the direction of eigenvectors being arbitrary, the standardization specifies the metrics only up to the sign: for example, any chosen eigenvector for economic complexity $\mathbf{x}$ is mathematically equivalent to any nonzero multiples $\gamma\mathbf{x}$, so that

$$\frac{\gamma\mathbf{x} - \text{mean}(\gamma\mathbf{x})}{\text{std}(\gamma\mathbf{x})} = \frac{\gamma}{|\gamma|}\,\frac{\mathbf{x} - \text{mean}(\mathbf{x})}{\text{std}(\mathbf{x})}. \tag{46}$$

Country Fitness and product Quality, on the other hand, are solutions to the following fix-point equations, solved recursively:

$$F_c = \frac{1}{\text{mean}\left(Q_p\right)} \sum_p M_{cp} Q_p, \tag{47}$$

$$Q_p = \frac{1}{\text{mean}(F_c)} \left[\sum_c M_{cp} \frac{1}{F_c}\right]^{-1}. \tag{48}$$

Normalizing the fixpoints yields the final indices:

$$\text{Fitness} = \frac{F}{\text{mean}(F)}, \tag{49}$$

$$\text{Quality} = \frac{Q}{\text{mean}(Q)}. \tag{50}$$